\documentclass[acmsmall, nonacm]{acmart}

\AtBeginDocument{%
  \providecommand\BibTeX{{%
    \normalfont B\kern-0.5em{\scshape i\kern-0.25em b}\kern-0.8em\TeX}}}

\setcopyright{acmlicensed}
\copyrightyear{2018}
\acmYear{2018}
\acmDOI{XXXXXXX.XXXXXXX}

\newtheorem{problem}{PROBLEM}

\acmConference[Conference acronym 'XX]{Make sure to enter the correct
  conference title from your rights confirmation emai}{June 03--05,
  2018}{Woodstock, NY}
\acmISBN{978-1-4503-XXXX-X/18/06}

\usepackage{graphicx}  
\usepackage{float}  
\usepackage{subfigure}  
\usepackage{subcaption}
\usepackage{multirow}
\usepackage{enumitem}
\usepackage{makecell}

\begin{document}

\title{Predicting Cascade Failures in Interdependent Urban Infrastructure Networks}
\author{Yinzhou Tang}
\affiliation{%
  \institution{Department of Electronic Engineering, Tsinghua University}
  \city{Beijing}
  \country{China}}
\email{tyz23@mails.tsinghua.edu.cn}

\author{Jinghua Piao}
\affiliation{%
  \institution{Department of Electronic Engineering, Tsinghua University}
  \city{Beijing}
  \country{China}}
\email{pjh22@mails.tsinghua.edu.cn}

\author{Huandong Wang}
\affiliation{%
  \institution{Department of Electronic Engineering, Tsinghua University}
  \city{Beijing}
  \country{China}}
\email{wanghuandong@tsinghua.edu.cn}

\author{Shaw Rajib}
\affiliation{%
  \institution{Graduate School of Media and Governance, Keio University}
  \city{Fujisawa}
  \country{Japan}}
\email{shaw@sfc.keio.ac.jp}

\author{Yong Li}
\affiliation{%
  \institution{Department of Electronic Engineering, Tsinghua University}
  \city{Beijing}
  \country{China}}
\email{liyong07@tsinghua.edu.cn}

\renewcommand{\shortauthors}{Yinzhou Tang, et al.}

\renewcommand{\shortauthors}{Yinzhou Tang, et al.}

\begin{abstract}
Cascading failures (CF) entail component breakdowns spreading through infrastructure networks, causing system-wide collapse. Predicting CFs is of great importance for infrastructure stability and urban function. Despite extensive research on CFs in single networks such as electricity and road networks, interdependencies among diverse infrastructures remain overlooked, and capturing intra-infrastructure CF dynamics amid complex evolutions poses challenges.
 To address these gaps, we introduce the \textbf{I}ntegrated \textbf{I}nterdependent \textbf{I}nfrastructure CF model ($I^3$), designed to capture CF dynamics both within and across infrastructures. $I^3$ employs a dual GAE with global pooling for intra-infrastructure dynamics and a heterogeneous graph for inter-infrastructure interactions. An initial node enhancement pre-training strategy mitigates GCN-induced over-smoothing.
 Experiments demonstrate $I^3$ achieves a 31.94\% in terms of AUC, 18.03\% in terms of Precision, 29.17\% in terms of Recall, 22.73\% in terms of F1-score boost in predicting infrastructure failures, and a 28.52\% reduction in terms of  RMSE for cascade volume forecasts compared to leading models. It accurately pinpoints phase transitions in interconnected and singular networks, rectifying biases in models tailored for singular networks. Access the code at https://github.com/tsinghua-fib-lab/Icube.

\end{abstract}





\maketitle

\section{Introduction}
\begin{figure}
  \centering
  \includegraphics[width=0.5\linewidth]{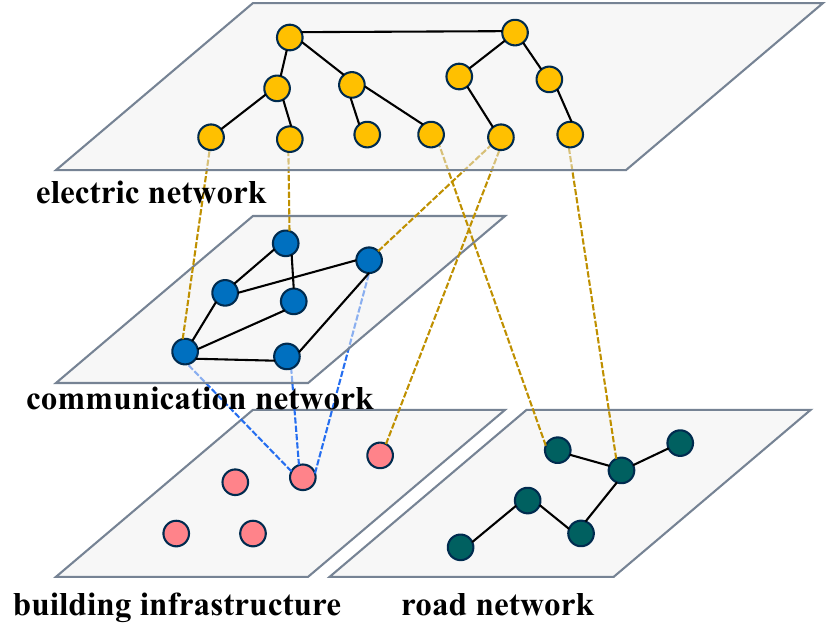}
  \caption{Illustration of interdependent urban infrastructure networks.}
  \label{urban_inf}
\end{figure}

Urban infrastructure networks refer to facilities related to the basic functioning of the city. These networks are not independent of each other. Instead, they can be regarded as the coupling of multiple infrastructure networks~\cite{mao2023detecting,buldyrev2010catastrophic}. Fig.~\ref{urban_inf} shows the relationship and connections between different urban infrastructures consisting of electric networks, road networks, communication networks, and building infrastructures. There are also various relationships between elements in different infrastructures, such as the connections in the same kind of infrastructure which are indicated by solid lines, and the connections between different kinds of infrastructure which are indicated by dashed lines. The former represents the relationship between different elements of the same infrastructure network, e.g., the power allocation relationship between different power plants in the electric network. The latter represents the relationships between elements in different infrastructure networks, e.g. the electric network affects the road network by supplying power to traffic lights, and the base station in communication networks provides signals to buildings.

Cascading Failure (CF) prediction is an important and widely studied problem in urban infrastructure networks. CF refers to the process in the failure starting from one or some elements causes the failure of other elements leading to a chain reaction of cascades in the infrastructure network~\cite{duan2019universal,buldyrev2010catastrophic}. It can be triggered by earthquakes, typhoons, or element failures due to aging of the network, \textcolor{black}{etc}. This process often causes catastrophic damage in cities, leading to the breakdown of city service and functionality~\cite{nerc200414,sun2019power}, and even posing a threat to the lives of citizens~\cite{mao2023detecting}. Accurate prediction of CF can help us understand key interconnections and vulnerabilities in infrastructure networks to improve system reliability~\cite{boyaci2022spatio,cassottana2022predicting}, or minimize the impact of infrastructure failures by early intervention~\cite{liu2021searching,ahmad2022prediction}, etc. \textcolor{black}{This is of great significance in ensuring the stable operation of urban infrastructure and the proper functioning of the city.} However, Predicting CF mainly faces the following three main challenges:

\begin{itemize}[]
\item[$\bullet$]\textbf{Dependencies between different kinds of infrastructures are diverse and complex~\cite{sturaro2018realistic,rahnamay2016cascading}.} CFs have been extensively studied in single networks, such as electric networks~\cite{liu2021searching,xie2020review}. However, real-world infrastructure networks are the combination of multiple infrastructure networks, i.e., interdependent networks. Different kinds of infrastructures have different dependencies with different mechanisms of coupling. This makes methods based on single infrastructure networks unable to capture the unique characteristics of interdependent networks. 

\item[$\bullet$]\textbf{Cascading failure dynamics within the same kind of infrastructure are affected by high-order and global dynamics~\cite{ghasemi2022higher}. }
Interactions in infrastructure networks are often not connected on a one-to-one basis. For example, for two connected nodes, the failure of one node does not necessarily mean that the other node will also fail.
In addition, there is a \textcolor{black}{high-order dynamic, which refers to the multi-hop dynamic in the network, and network resilience} during cascading failures.
Moreover, failures in infrastructure networks are not only affected by other elements in \textcolor{black}{the same} network but also by the coupled network. This complex cascading failure dynamics within the same kind of infrastructure poses a challenge to the predictions of CF.

\item[$\bullet$]\textbf{The complex coupling dependency between initial and final failed nodes is difficult to model.} In urban infrastructure CFs, the number of positive and negative samples of initially failed labels is imbalanced, which makes it difficult to capture the relationship between them~\cite{varbella2023geometric,qu2019gmnn}. There is a significant locality of CF propagation in infrastructure networks, especially before the phase transition, when CF only propagates locally rather than causing a global failure~\cite{ash2007optimizing,xing2020cascading}. It indicates a high local structural dependency between labels of connected nodes~\cite{hua2022graph,jin2021survey}. Existing methods are mainly based on modeling the node feature propagation while ignoring the local structural dependencies between node labels, which leads to imprecise predictions for CF~\cite{zhou2022graph}.
\end{itemize}

Based on the above challenges, in this paper, we propose an \textbf{I}ntegrated \textbf{I}nter-dependent \textbf{I}nfrastructure cascade failure model ($I^3$) to capture not only the dynamics of single but also interdependent networks and the coupling mechanisms to predict the CF in urban infrastructures. The model is based on a double GAE backbone and we designed a global pooling pre-train task to capture the dynamics within one infrastructure. Then, we used a heterogeneous graph structure to capture the dynamics between different kinds of infrastructure to model the unique characteristics in interdependent networks. We also add an initial node enhancement pre-training task based on the idea of label propagation~\cite{iscen2019label} to model the complex coupling dependency between the initial and final failed nodes.

Our contribution can be summarized as follows:
\begin{itemize}[]
\item[$\bullet$]We propose $I^3$, an urban infrastructure CF prediction model that captures the dynamics of CFs within the same kind of infrastructure and between different kinds of infrastructure.
\item[$\bullet$]We apply the idea of label propagation to CF prediction by designing an initial node enhancement module to avoid the over-smoothing problem that occurs during the prediction process using GCN-based models.
\item[$\bullet$]Extensive experiments demonstrate that comparing the state-of-the-art baselines, $I^3$ can achieve a 31.94\% improvement in terms of AUC, 18.03\% in terms of precision, 29.17\% in terms of recall, and 22.73\% in terms of F1-score. For CF volume, it can achieve a 28.52\% improvement in RMSE. We also verify through a case study that $I^3$ can correctly predict the phase transition of network breakdown in infrastructure networks and can effectively predict CFs in different networks.
\end{itemize}

\section{Related Works}
Existing CF prediction methods can be categorized into methods based on single networks and interdependent networks.

\subsection{CF prediction on single network}

Most of the existing work on CF prediction in urban infrastructure networks focuses on single networks, e.g., given an initially destroyed power plant, predicting the power plants that will be destroyed afterward. These methods are often based on expert knowledge in the domain to propose CF prediction algorithms based on some rule that is presented in the form of differential equations. For example, Dobson et al. proposed OPA~\cite{dobson2001initial} based on DC(Direct Current) flow calculation analysis, which enables cascading failure modeling and analysis of the grid, and Riot et al. proposed the Manchester model~\cite{rios2002value} to address the lack of AC(Alternating Current) features in the model by focusing on AC flows. Thereafter, a large number of DC and AC models have been proposed to model cascading failures in electric networks~\cite{mei2009improved,ciapessoni2013cascading}. 

Recently, more and more approaches have begun to predict cascading failures in single networks through machine learning models. For example, Nakarmi et al used the Markov chain model to predict the cascade size distribution based on the location of initial faults in each community identified within the power system~\cite{nakarmi2020markov}. Zhu et al proposed a physically informative GNN based on a data-driven model to solve the tidal equations so that the output satisfies the laws of physics and makes the results more physically interpretable~\cite{zhu2022cascading}. Basak et al designed a model based on LSTM that can capture spatial information and interconnection of transportation networks and predict cascading failures~\cite{basak2019data}.

Although CF prediction algorithms on single networks can better take into account the features within the network, they are unable to capture the impact of other kinds of infrastructures on the current network, which makes the prediction results suboptimal.

\subsection{CF prediction on interdependent network}

Real-world infrastructure networks are not isolated and elements of different networks are interdependent. If a small failure occurs in one network, it can cause components of other networks to fail and propagate, leading to the collapse of the entire system~\cite{valdez2020cascading}. Thus, more and more attention is being paid to the prediction of CFs on interdependent networks.

Buldyrev et al developed a simple model of interdependence between networks and showed that such systems might suddenly collapse under random faults~\cite{parandehgheibi2014mitigating}. Parandehgheibi et al predicted CF on an electric network-communication network and proposed a two-stage control strategy to mitigate cascading faults. Sturaro et al focused on the interdependence between grid-communication networks and proposed the HINT model for CF prediction~\cite{sturaro2018realistic}.

There are also many data-driven methods for predicting CF in interdependent networks. Cassottana et al. developed a resilience prediction model using the infrastructure simulation model ML algorithm for CF prediction in water-grid-transportation networks~\cite{cassottana2022predicting}. Rahnamay-Naeini et al. used an interdependent Markov chain framework to capture the interdependence between two critical infrastructures to predict interdependent networks in CF~\cite{rahnamay2016cascading}. Mao et al. developed a graph neural network system based on reinforcement learning to detect vulnerable nodes in infrastructure networks~\cite{mao2023detecting}.

These coupled network-based approaches either model only part of the network, e.g. not considering the communication network or the road network, resulting in them being one-sided, or do not use a graph structure to extract the relationships between different infrastructures.

\section{Problem Formulation}

In this paper, we aim to predict the set of nodes that will eventually fail on an infrastructure network given the initial failed nodes. 
We consider an interdependent network with four types of nodes and model it as a heterogeneous graph structure. Formally, the infrastructure network $G(V,E)$ consists of a series of single networks $G_{sg}(V_{sg}, E_{sg})$ including an electric network $G_{e}(V_{e},E_{e})$, a road network $G_{r}(V_{r},E_{r})$, a communication network $G_{c}(V_{c},E_{c})$, and building infrastructure $G_{b}(V_{b},E_{b})$. $V=\{ V_{e} \bigcup V_{r} \bigcup V_{c} \bigcup V_{b}\}$ denotes all the elements in the infrastructure network, and the node attribute of each node $n$ represents its state, with 0 denoting failure and 1 denoting normal. $E=\{ E_{e} \bigcup E_{r} \bigcup E_{c} \bigcup E_{b} \bigcup E_{cp} \}$ denotes the relationship in the same kind of infrastructure and the dependencies between different infrastructures $E_{cp}$. Based on the above notation, we can define the CF prediction problem as follows:

\begin{problem}[Cascade Failure Prediction]
Given an infrastructure network heterogeneous graph $G(V, E)$ and an initial set of damaged nodes $D$ under a certain CF case, we aim to predict the set of nodes that will eventually fail $D'$.
\end{problem}

\begin{figure*}
  \centering
  \includegraphics[width=0.99\linewidth]{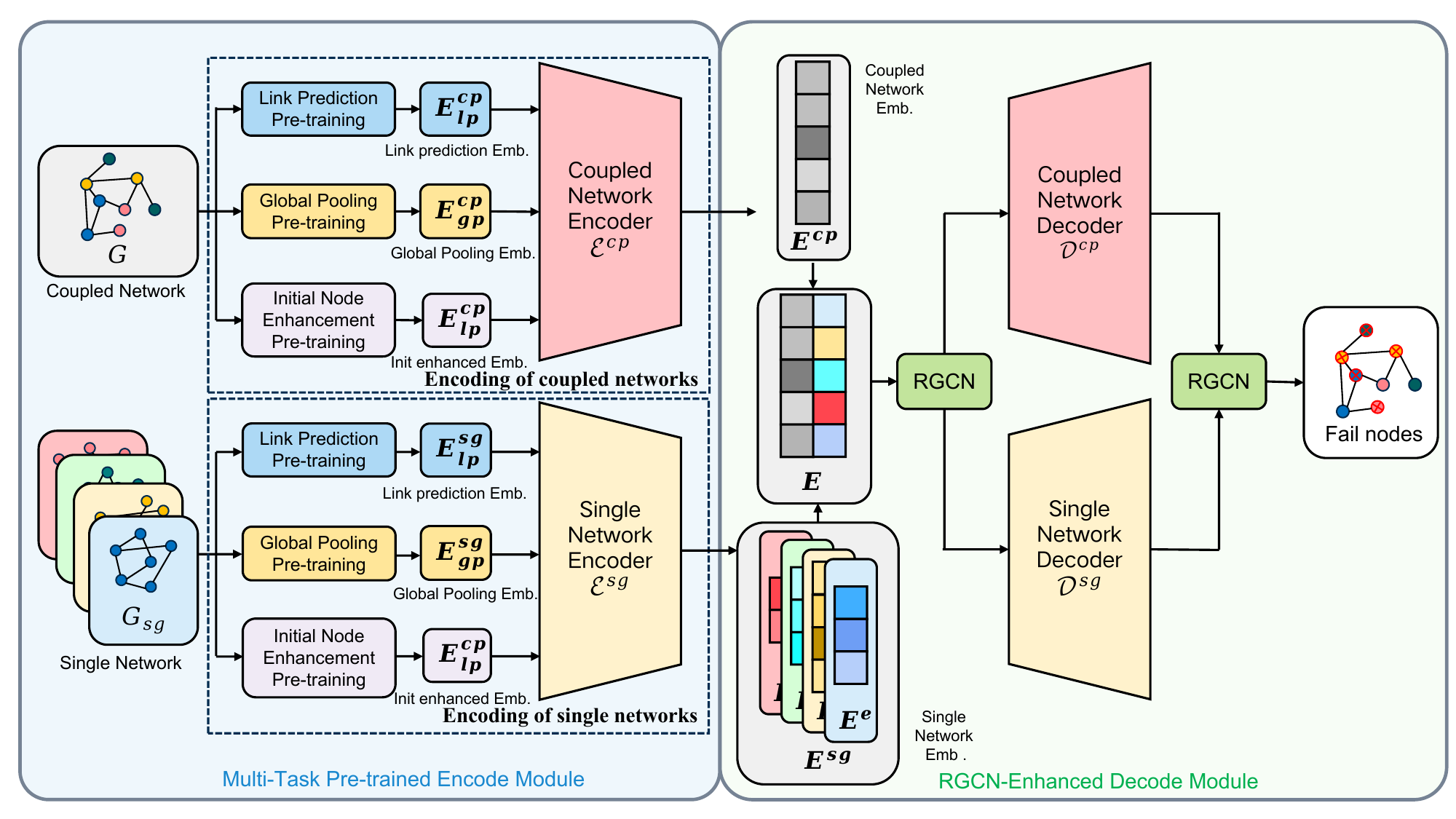}
  \caption{Model architecture of $I^3$.}
  \label{main}
\end{figure*}

\section{Method}

In order to address the three challenges presented in Section~1, we designed an \textbf{I}ntegrated \textbf{I}nterdependent \textbf{I}nfrastructure Cascade Failure Model ($I^3$) to predict CFs in infrastructure networks. The structure of $I^3$ is shown in Fig.~\ref{main}. \textcolor{black}{We used a double GAE backbone consisting of a multi-task pre-trained encode module and an RGCN-enhanced decode module to extract the features of coupled networks and single networks, respectively.} $I^3$ first encodes the features in different networks. During the encoding process, we designed three different pre-training tasks to get three embeddings. \textcolor{black}{We use the embedding of link prediction pre-training $\mathbf{E_{lp}^{cp}}$ and $\mathbf{E_{lp}^{sg}}$ to capture the topology of the graph, the embedding of global pooling pre-training $\mathbf{E_{gp}^{cp}}$ and $\mathbf{E_{gp}^{sg}}$ to extract the high-order dynamics and global features, respectively, and an initial node enhancement module was introduced to get the initial node enhanced embedding $\mathbf{E_{ie}^{cp}}$ and $\mathbf{E_{ie}^{sg}}$ to avoid over-smoothing in GCN. }\textcolor{black}{In the above description, the superscript $s$ denotes the embedding for single networks and $c$ denotes the embedding for the coupled network. }Then $I^3$ predicts the propagation of CF and decodes them to get the final failed node set in the decode module. \textcolor{black}{In the following part, we will introduce the multi-task pre-trained encode module and three pre-training tasks to get three embeddings in Section 4.1 and the RGCN-enhanced decode module in Section 4.2.}

\subsection{Multi-Task Pre-trained Encode Module}
In this section, we will explain how to extract different features of different networks and encode them. Specifically, we use two encoders: Coupled Network Encoder $\mathcal{E}^{cp}$ and Single Network Encoder $\mathcal{E}^{sg}$. They are respectively responsible for encoding the features of Coupled Network $G$ and Single Network $G_{sg}$ to get Coupled Network Embedding $\mathbf{E^{cp}}$ and Single Network Embedding $\mathbf{E^{sg}}$. For $\mathcal{E}^{sg}$, there is only one instantiation while there are four instantiations for $\mathcal{E}^{cp}$ concerning different single networks. Each instantiation for $\mathcal{E}^{sg}$ represents one kind of urban infrastructure network. For each encoder, we design three different pre-train tasks to get different embeddings.



\subsubsection{Link Prediction Pre-training}
The propagation of CF is affected by the topology of infrastructure networks. In order to capture the graph topology, we follow~\cite{mao2023detecting} and apply a link prediction pre-training task. This is because the topology of a network implicitly reveals the indications of similarity and correlation among nodes, which can be used as supervised signals to obtain the features of different networks. Additionally, there are cascade relationships between infrastructures that are not directly connected in the urban network. Consequently, the task can be considered a link prediction problem~\cite{grover2016node2vec}. Its goal is to estimate the probability of the edge existence from node attributes and the observed edges. Specifically, we build a positive graph $G_p$ by taking all the edges present in the graph as a positive set and sample the edges that are not present as a negative set to form a negative graph $G_n$, and the link prediction score $S$ can be defined as follows:
\begin{align}
	S(\mathbf{E_{lp}},G)=\{ z_i^T \cdot z_j, \forall e_{ij} \in E \},	\label{link prediction}
\end{align}
in which $z_i$ and $z_j$ refers to the embedding for node $i$ and $j$ in graph $G$, and $\mathbf{E_{lp}}$ refers to the node embedding of the nodes. It computes the inner product of the embedding vectors of the two nodes on each existing edge on a given graph $G$ as the prediction weights. 
Then we use the margin loss function to get the representation of the nodes as follows:

\begin{align}
	\mathcal{L}_{lp}=\operatorname{MEAN}(\max(0,M-S(\mathbf{E_{lp}},G^p)+S(\mathbf{E_{lp}},G^n)))+\lambda||\Theta||^2,	\label{link prediction loss}
\end{align}
where $M$ is a constant parameter and $\lambda||\Theta||^2$ is a $L_2$ regularization term.

By pre-training on different graphs, for each node we can get an embedding with respect to the similarity between nodes, which we use as a link prediction embedding $\mathbf{E_{lp}^{cp}} \in \mathbf{R}^{|V| \times F_{lp}^D}$ and $\mathbf{E_{lp}^{sg}} \in \mathbf{R}^{|V| \times F_{lp}^D}$ and $F_{lp}^D$ refers to the dimension of node attribute.


\subsubsection{Global Pooling Pre-training}
As presented in Section~1, CF dynamics in infrastructure networks are not only affected by local topology but also by higher-order dynamics such as multi-hop interactions and global evolution. Thus, we need to get another embedding that models the global features. We design a global pooling pre-training task to extract the global pooling embedding $\mathbf{E_{gp}^{cp}}$ and $\mathbf{E_{gp}^{sg}}$. Specifically, we employ a DiffPool network~\cite{ying2018hierarchical} to perform a regression task for end-to-end pre-training to extract global features.

In order to extract the global information, we build a new graph $G_{D}$ with the same graph structure as the input graph, but with different node attributes. In $G_{D}$, the node attributes of the heterogeneous graph are the state of the nodes in the initial case and the shortest path length of the current node from the initial failed nodes. We conduct a regression task on the DiffPool with $G_{D}$ as the input and the prediction target is the average of the shortest distances of all nodes in the graph from the initial failed nodes $s_G$, which can be formulated as:
\begin{align}
	s_{G}=\frac{\sum_{n \in V}{L_{ni}}}{|V|}, \label{s}
\end{align}
in which $L_{ni}$ refers to the shortest distance between node $n$ and node $i$, node $i$ refers to the initial failed node. $|V|$ refers to the number of nodes in the graph. When there is more than one initial failed node, the $s_{G}$ can be formulated as:
\begin{align}
	s_{G}=\frac{\sum_{i \in D}\sum_{n \in V}{L_{ni}}}{|V|},  \label{s1}
\end{align}
in which $D$ refers to the initial failed node set. Afterward, we stack $L$ layers of GNNs, with the $l$th layer applying the embedding of the $l-1$th layer after pooling and learning how to assign nodes to the clusters in the $l$th layer.

We define $S^{(l)}$ as the cluster assign matrix learned at layer $l$. Then the embedding of each cluster after allocation can be obtained by the following equation:
\begin{align}
	X^{(l+1)}=S^{(l)T}Z^{(l)} \in \mathbb{R}^{b_{l+1} \times d},	\label{X}
\end{align}
in which $b_{l+1}$ refers to the cluster number in $l+1^{th}$ layer and d refers to the feature dimension. And the adjacency matrix between the new clusters can be generated by the following equation:
\begin{align}
	A^{(l+1)}=S^{(l) T} A^{(l)} S^{(l)} \in \mathbb{R}^{b_{l+1} \times b_{l+1}}. \label{A}
\end{align}

In the DiffPool, we use two GNNs to generate the feature matrix and the allocation matrix of the $l^{th}$ layer, respectively, which can be expressed as:
\begin{align}
	Z^{(l)}=\mathrm{GNN}_{l, \text { embed }}\left(A^{(l)}, X^{(l)}\right), \label{Z}
\end{align}
\begin{align}
	S^{(l)}=\operatorname{softmax}\left(\operatorname{GNN}_{l, \text { pool }}\left(A^{(l)}, X^{(l)}\right)\right). \label{S}
\end{align}
The loss function we use is the RMSE of the predicted and labeled average of the shortest distance of all nodes from the initial node, which can be formulated as:
\begin{align}
	\mathcal{L}_p=RMSE(s_G,\hat{s_G}). \label{Lossp}
\end{align}

After pre-training, the well-trained DiffPool model should be able to capture the global features of the network given the initial failed nodes. To better capture the higher-order interactions at different levels, we construct the Global Pooling embedding $\mathbf{E_{gp}^{cp}}$ and $\mathbf{E_{gp}^{sg}} \in \mathbb{R}^{N \times (L-1)}$, in which $N$ refers to the number of nodes in the network. The $l^{th}$ element of the $n^{th}$ row $e_p^{n,l}$ denotes a difference formulated as:
\begin{align}
	e_p^{n,l}=s_n^l-s_G. \label{e}
\end{align}
in which $s_n^l$ refers to the average shortest path length of the cluster to which the $n^{th}$ node belongs at the $l^{th}$ level of the DiffPool.

From this, we can obtain embedding representations with high-order dynamics at different levels and global dynamics for coupled networks and single networks, respectively.

\subsubsection{Initial Node Enhancement Pre-training}
In urban infrastructure CFs, there are two sets of labels: the initial failed labels and the final failed labels. However, the positive and negative samples of the initial failed nodes are imbalanced, which indicates that the number of failed nodes in the initial state is much less than the number of normal nodes~\cite{hua2022graph,zhou2022graph}. What's more, the propagation of CF is of high locality, especially before the CF phase transition~\cite{ash2007optimizing,xing2020cascading}. In this case, CF only propagates locally instead of leading to a global infrastructure failure. For example, the neighbors of the initial failed nodes are more likely to fail. This indicates that CFs have a great local structural dependency between connected nodes~\cite{jin2021survey,iscen2019label}. In order to capture the complex dependency between initial and final failed nodes, we introduce another semi-supervised pre-training task, GMNN~\cite{qu2019gmnn}, which is a method for fusing node label correlations based on the idea of label propagation. Intuitively, GMNN uses the paradigm of label propagation instead of feature propagation, which follows the insight of closer nodes are more likely to share the same label~\cite{iscen2019label}.

Specifically, GMNN uses CRF to model the joint distribution between labels, which is optimized by using the pseudo-likelihood variational EM algorithm. In this case, in M-step a GNN is used to model the dependencies between labels, and in E-step another GNN is used to learn the feature representation of nodes to predict label attributes.

In the M-step, it can be regarded as a learning procedure to annotate unlabeled nodes with $q_\theta$. We set $\hat{y_V}=(y_L,\hat{y_U})$ and update $p_\phi$ with the following formula:
\begin{equation}
	O_\phi=\sum_{n \in V} \log p_\phi\left(\hat{\mathbf{y}}_n \mid \hat{\mathbf{y}}_{\mathrm{NB}(\mathrm{n})}, \mathbf{x}_V\right), \label{Op}
\end{equation}
in which $x_V$ refers to the node attribute and we can use a GNN to parameterize $ \log p_\phi\left(\hat{\mathbf{y}}_n \mid \hat{\mathbf{y}}_{\mathrm{NB}(\mathrm{n})}, \mathbf{x}_V\right)$ by:
\begin{equation}
	p_\phi\left(\mathbf{y}_n \mid \mathbf{y}_{\mathrm{NB}(n)}, \mathbf{x}_V\right)=\operatorname{Concat}\left(\mathbf{y}_n \mid \operatorname{softmax}\left(W_\phi \mathbf{h}_{\phi, n}\right)\right), \label{p}
\end{equation}
in which $W_\phi$ refers to the GNN embedding, $\mathbf{h}_{\phi, n}$ refers to the node embedding and $NB(n)$ is the neighbor of node $n$. 

The E-step is an inference procedure to annotate unlabeled objects with $p_\phi$ and $\hat{y_V}$. We update $q_\theta$ with 

\begin{equation}
	O_\theta=\sum_{n \in U} \mathbb{E}_{p_\phi}\left[\log q_\theta\left(\mathbf{y}_n \mid \mathbf{x}_V\right)\right]+\sum_{n \in L} \log q_\theta\left(\mathbf{y}_n \mid \mathbf{x}_V\right), \label{Oq}
\end{equation}
where $q_\theta$ can also be implemented by a GNN formulated as:
\begin{equation}
	q_\theta\left(\mathbf{y}_n \mid \mathbf{x}_V\right)=\operatorname{Concat}\left(\mathbf{y}_n \mid \operatorname{softmax}\left(W_\theta \mathbf{h}_{\theta, n}\right)\right), \label{q}
\end{equation}
where $W_\theta$ refers to the GNN embedding and $\mathbf{h}_{\theta, n}$ refers to the node embedding.

With the above E-step and M-step one can alternatively optimize the \textbf{E}vidence \textbf{L}ower \textbf{BO}unds (ELBO), which can be formulated as:
\begin{equation}
	\mathcal{L}_{ie}=\mathbb{E}_{q_\theta\left( y_U | x_V \right)}\left[{\log p_\phi \left({y_L,y_U | x_V}\right)-\log q_\theta\left({y_U | x_V}\right)}\right], \label{q}
\end{equation}
where $\hat{y}_V$ denotes the predicted label of the set $V$ of all nodes, $y_L$ denotes the label of the node in the set of existing labels, and $\hat{y}_U$ denotes the predicted label of the node in the set of unknown labels.

Therefore, we use the coupled and single infrastructure networks in $I^3$, where the nodes are characterized by the state of the node at the initial moment, and we use the GMNN for semi-supervised propagation. Then the hidden layer embedding of the GNN of the last training $q_\theta$ is extracted as $\mathbf{E_{ie}^{cp}}$ and $\mathbf{E_{ie}^{sg}}$. Thus, the embedding of each node contains the information of the initial failed node.

\subsubsection{Graph Feature Encoding}
After obtaining the three embedding for the coupled and single networks respectively, we concatenate them to obtain the embedding of the coupled and single graphs through two GAE encoders, which can be represented as follows:
\begin{equation}
	\mathbf{E^{cp}}=\mathcal{E}^{cp} \left( \text{Concat}\left(\mathbf{E^{cp}_{lp}},\mathbf{E^{cp}_{gp}},\mathbf{E^{cp}_{ie}}\right) \right),\label{encoder1}
 \end{equation}
 \begin{equation}
    \mathbf{E^{sg}}=\mathcal{E}^{sg} \left( \text{Concat}\left(\mathbf{E^{sg}_{lp}},\mathbf{E^{sg}_{gp}},\mathbf{E^{sg}_{ie}}\right) \right). \label{encoder2}
\end{equation}
It is worth mentioning that there are four different instantiations for the Single Network Encoder $\mathcal{E}^{sg}$. For example, we can get the embedding of the electric network by conducting Eq~\ref{encoder2} on the electric network. Thus, we can use Eq~\ref{encoder1} and Eq~\ref{encoder2} to predict the propagation of CF.

\subsection{RGCN-Enhanced Decode Module}
After obtaining the encoded coupled network embeddings and single network embeddings, we can predict the CF and decode them by RGCN-enhanced decode module. Owing to the fact that real-world urban infrastructure networks are jointly coupled with multiple different infrastructure networks, it is important to distinguish different infrastructures when modeling the CF propagation. Therefore, we choose to use \textbf{R}elational \textbf{G}raph \textbf{C}onvolutional \textbf{N}etworks (RGCN) to model different kinds of infrastructures. Specifically, we first concatenate the coupled network embedding and single network embedding to obtain a representation of each infrastructure, which can be represented as follows:
\begin{equation}
    E_i=\left\{\begin{array}{l}
\text { Concat }\left(E_i^{c p}, E_i^e\right), \text { if } i \in V_e \\
\text { Concat }\left(E_i^{c p}, E_i^r\right), \text { if } i \in V_r \\
\text { Concat }\left(E_i^{c p}, E_i^c\right), \text { if } i \in V_c \\
\operatorname{Concat}\left(E_i^{c p}, E_i^b\right), \text { if } i \in V_b
\end{array}\right. 
\end{equation}
in which $E_i$ refers to the node embedding of node $i$, $E_i^{cp}$ refers to the coupled network embedding of node $i$ and $E_i^{e}$, $E_i^{r}$, $E_i^{c}$, $E_i^{b}$ refers to the single network embedding of node $i$ in electric network, road network, communication network and building network, respectively.

Since we are only concerned about the set of nodes that failed in the end for the CF propagation process, we use an RGCN to predict with the following equation:
\begin{equation}
	\mathbf{\hat{E}}=RGCN(\mathbf{E},G). \label{predict E}
\end{equation}

RGCN is a graph convolutional network used on heterogeneous graphs. For each node, the aggregation process of RGCN considers the contribution of different types of edges and neighbors to the node. For each node $i$, its embedding can be expressed as:
\begin{equation}
	h_i^{(l+1)}=\sigma\left( \sum_{r \in R} \sum_{j \in N_i^r} \frac{1}{c_{i,r}}W_r^{(l)}h_j^{(l)}+W_0^{(l)}h_i^{(l)} \right), \label{RGCN}
\end{equation}
in which $h_i^{(l+1)}$ refers to the embedding of node $i$ in $l+1$th layer, $r$ refers to a certain type of edge, $N_i^r$ refers to the set of neighbors of node $i$ with edge type $r$ and $c_{i,r}$ refers to the normalize parameter. An edge type $r$ includes homogeneous such as the edge between two power stations, and heterogeneous edges such as the edge between a base station and a building.

After the RGCN, the embedding of nodes contains information about different types of nodes and edges, which can be used to model the propagation of CF in heterogeneous networks. Afterward, we need to decode the embedding of the nodes. Specifically, we use two decoders, a coupled network decoder and a single network decoder. After decoding, the obtained features need to be concatenated and go through an RGCN to get the final probability of the failure of each node. This process can be represented as:
\begin{equation}
	\mathbf{\hat{X}}=RGCN(\operatorname{Concat}(\mathcal{D}^{cp}(\mathbf{\hat{E}}),\mathcal{D}^{sg}(\mathbf{\hat{E}})),G). \label{predict node}
\end{equation}
After obtaining the predicted results, we compute the cross-entropy loss with the labeled ground truth and then perform back-propagation for gradient optimization in order to train the model, and the expression of the loss function is as follows:
\begin{equation}
	\mathcal{L}=\operatorname{CrossEntropy}(\mathbf{X},\mathbf{\hat{X}}), \label{loss}
\end{equation}
where $\mathbf{X}$ refers to the ground truth and $\mathbf{\hat{X}}$ refers to the predict fail possibility.

\begin{table}
\renewcommand\arraystretch{0.99}
    \centering
	\caption{
       Summary of the homogeneous elements used in the synthetic and real-world network.}
	\begin{tabular}{l||c|c|c|c|c|c}
    	\hline
         {Dataset} & {$\#N_{elec}$} & {$\#N_{road}$} & {$\#N_{com}$}& {$\#E_{elec}$} & {$\#E_{road}$} & {$\#E_{com}$}\\ 
        \cline{1-7}
        {Synthetic}&{10,227} & {4,825}  & {20,229} & {7,799} & {20,352}  & {44,282} \\
        \cline{1-7}
        {Real-world}& {684} & {7,190}  & {12,992} & {548} & {3,845}  & {6,016}\\
         \hline
	\end{tabular} 
    \label{tab:dataset1}
\end{table}

\begin{table}
\renewcommand\arraystretch{0.99}
    \centering
	\caption{
       Summary of the heterogeneous elements used in the synthetic and real-world network. The backslash indicates that there is no such data.}
	\begin{tabular}{l||c|c|c|c}
    	\hline
        {Dataset} & {$\#E_{elec-road}$} & {$\#E_{elec-com}$} & {$\#E_{elec-AOI}$} & {\#$E_{com-AOI}$} \\ 
        \cline{1-5}
         {Synthetic}&{7,799} & {20,352}  & {21,569} & {37,279}  \\
         \cline{1-5}
         {Real-world}&{6,496} & {7,190}  & {$/$} & {$/$}  \\
         \hline
	\end{tabular} 
    \label{tab:dataset2}
\end{table}


\begin{figure*} 
	\centering  
	\subfigtopskip=2pt 
	\subfigbottomskip=2pt 
	\subfigcapskip=-5pt 
	\subfigure[Difference between $I^3$ predicted heat map and the ground truth.]{
		\label{Exp3-Pred}
		\includegraphics[width=0.4\linewidth]{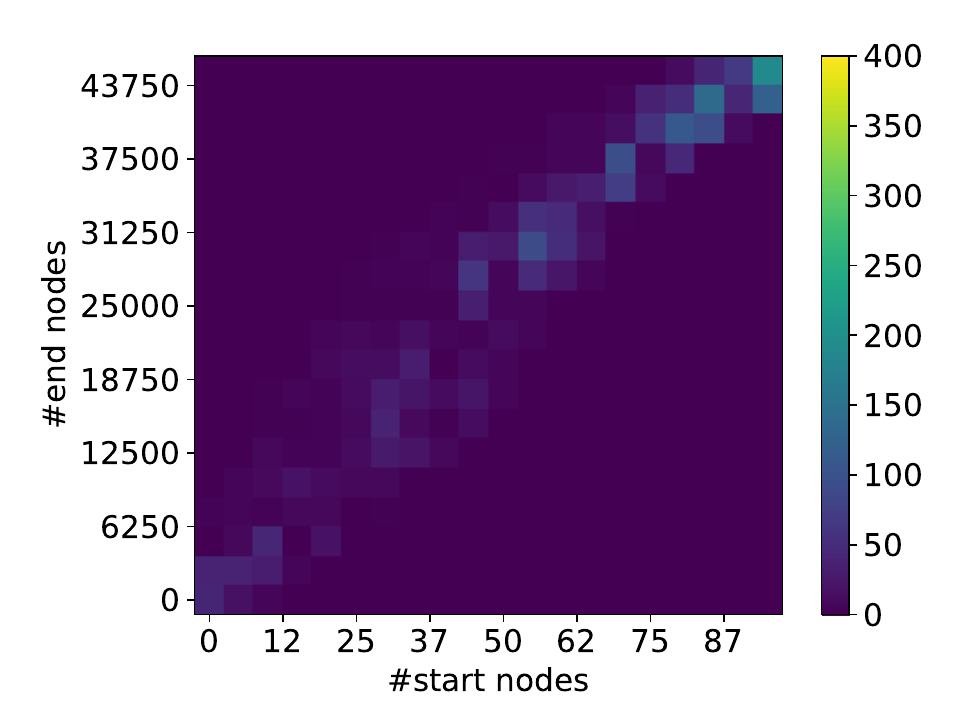}}
	\subfigure[Difference between HINT predicted heat map and the ground truth.]{
		\label{Exp3-Baseline}
		\includegraphics[width=0.4\linewidth]{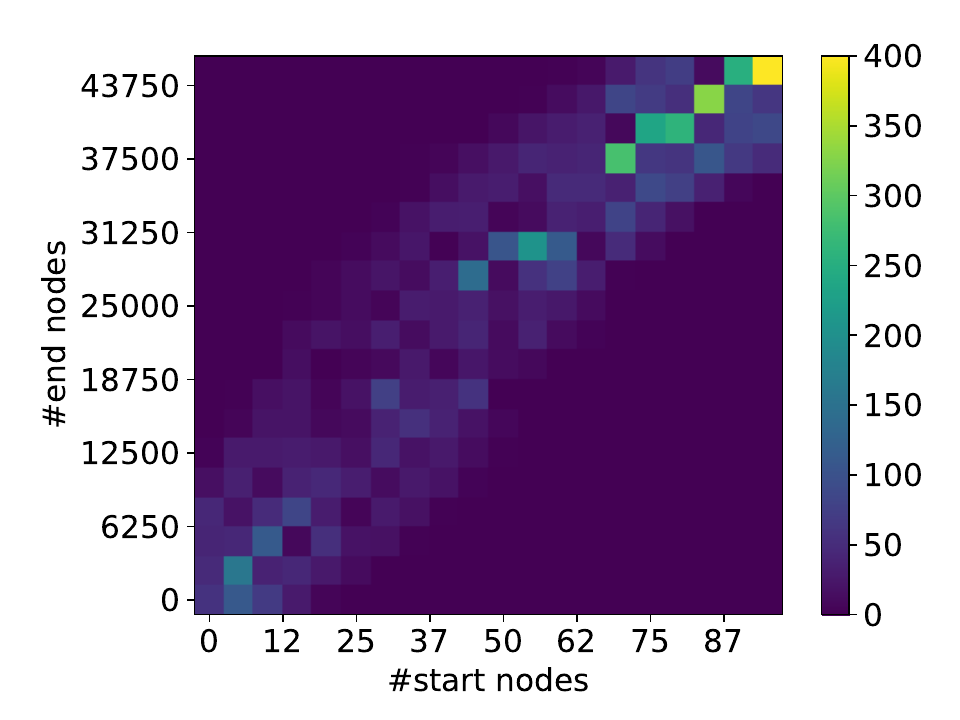}}
	\caption{Difference between predicted heat map of different models and the ground truth.}
	\label{Exp3}
\end{figure*}

\section{Experiments}
In this section, we aim to explore the performance of our model in several aspects and answer the following questions:
\begin{itemize}[]
\item[$\bullet$]RQ1: Does our model achieve the best performance among other CF prediction models?
\item[$\bullet$]RQ2: Are each of our three proposed embedding able to model different features of infrastructure networks?
\item[$\bullet$]RQ3: Does our model achieve accurate prediction even with different combinations of initial damages?
\item[$\bullet$]RQ4: Is our model able to predict the phase transition point of failure in infrastructure networks?
\item[$\bullet$]RQ5: Is our model able to capture different collapses and phase transition points in different single networks?
\end{itemize}

\subsection{Datasets}
\textcolor{black}{In our experiment, we build two datasets to evaluate our model, a synthetic dataset and a real-world dataset. The specific nodes and connection relationships are shown in Table~\ref{tab:dataset1} and Table~\ref{tab:dataset2}.} 
\subsubsection{Synthetic Dataset}
We build a synthetic dataset based on the urban simulator proposed in~\cite{zhang2022mirage}. 
The synthetic interdependent network $G_{inter}$ consists of three networks: electric network $G_{elec}$, road network $G_{road}$, and communication network $G_{com}$. We constructed a heterogeneous graph structure according to the types of nodes and edges. 
In the electric network, it consists of different levels of power plants such as 500kV and 220kV. In addition, we also consider modelling the functions of an AOI(area of interest), which refers to a collection of one or more buildings. For the electric and communication networks, we added edges between them and the AOI within the network to model the electric and communication function of the AOI. For the number of initial failed nodes, we choose 0-20 initial failed nodes, and given the initial failed node $i$, we randomly select $i$ nodes from the network as the initial failed nodes and use~\cite{zhang2022mirage} to predict the propagation of CF in the network to get the set of final failed nodes. For each $i$, we construct 100 cascade records. 
\subsubsection{Real-world Dataset}
\textcolor{black}{We also built a real-world dataset consisting of a electric network $G_{elec}$, a road network $G_{road}$ and a communication network $G_{com}$. The real-world topology is based on the electric network in Bonneville Power Administration Data\footnote{https://data-bpagis.hub.arcgis.com}, communication network in OpenCellid dataset\footnote{https://opencellid.org}, road network in North America from the Bureau of Transportation Statistics\footnote{https://hub.arcgis.com/datasets/usdot::north-american-roads/about}. We also derived 684 cascade failure records are derived from North America Bonneville Power Administration Dataset\footnote{https://transmission.bpa.gov/Business/Operations/Outages}. The data and a more detailed process of building the real-world dataset is available at https://github.com/tsinghua-fib-lab/Icube.}

\subsection{Experimental Settings}
\subsubsection{Dataset Splitting}
For both datasets, we choose 60\% as the training set, 20\% as the validation set, and 20\% as the testing set. \textcolor{black}{We also conduct a 5-fold cross-validation on every dataset and baselines.}
\subsubsection{Metrics}
In this experiment, we measure the accuracy of prediction at the node level and volume level, respectively.
For the accuracy of node prediction, we use AUC, Precision, Recall, and F1-score as prediction metrics with the purpose of detecting whether every node is correctly predicted. The formula for calculating the metrics can be represented as follows:
\begin{align}
	\text{Pre}=\frac{TP}{TP+FP},\\
    \text{Rec}=\frac{TP}{TP+FN},\\
    \text{F1}=\frac{2\times\text{Pre}\times\text{Rec}}{\text{Pre}+\text{Rec}},\label{Metric}
\end{align}
in which $TP$, $FP$, and $FN$ refer to the number of true positive, false positive, and true negative samples. For the accuracy of volume prediction, we use RMSE as a prediction metric, which aims to measure the accuracy of predicting the number of failed nodes. The formula is as follows:
\begin{align}
	\text{RMSE} = \sqrt{\frac{1}{n} \sum_{i=1}^{n} (y_i - \hat{y}_i)^2},\label{RMSE}
\end{align}
in which $n$ denotes the number of samples, $y_i$ and $\hat{y_i}$ refer to the true value and the model predicted value of the cascade volume, respectively.
\begin{table*}
\renewcommand\arraystretch{0.99}
    \centering
    \tiny
	\caption{
       CF prediction performance evaluation results on interdependent networks in \textbf{SYNTHETIC} dataset. The backslash indicates that the baseline cannot be applied to the analysis of the network. The best results are highlighted in bold.}

	\begin{tabular}{l||c|c|c|c||c||c|c||c|c||c|c}
    	\hline
        \multirow{3}{*}{} & \multicolumn{5}{c||}{$G$}& \multicolumn{2}{c||}{$G_{e}$} & \multicolumn{2}{c||}{$G_{r}$}  & \multicolumn{2}{c}{  $G_{c}$}    \\ \hline
        & Pre & Rec & F1 & AUC & RMSE& F1& RMSE& F1& RMSE& F1& RMSE \\ \hline
        ICM~\cite{goldenberg2001talk}  & \makecell[c]{0.05\\[-0.75ex]$\pm$0.03} & \makecell[c]{0.17\\[-0.75ex]$\pm$0.05} & \makecell[c]{0.07\\[-0.75ex]$\pm$0.04} & \makecell[c]{0.52\\[-0.75ex]$\pm$0.05} & \makecell[c]{6542.98\\[-0.75ex]$\pm$12.58}& \makecell[c]{0.08\\[-0.75ex]$\pm$0.05} & \makecell[c]{6647.99\\[-0.75ex]$\pm$13.64}& \makecell[c]{0.09\\[-0.75ex]$\pm$0.06} & \makecell[c]{6427.99\\[-0.75ex]$\pm$15.42}& \makecell[c]{0.06\\[-0.75ex]$\pm$0.05} & \makecell[c]{6544.71\\[-0.75ex]$\pm$13.22} \\
    	\cline{1-12}
        GCN~\cite{kipf2016semi}  & \makecell[c]{0.06\\[-0.75ex]$\pm$0.04} & \makecell[c]{0.19\\[-0.75ex]$\pm$0.08} & \makecell[c]{0.09\\[-0.75ex]$\pm$0.04} & \makecell[c]{0.52\\[-0.75ex]$\pm$0.07} & \makecell[c]{6127.85\\[-0.75ex]$\pm$32.54}& \makecell[c]{0.09\\[-0.75ex]$\pm$0.04} & \makecell[c]{6138.74\\[-0.75ex]$\pm$33.65}& \makecell[c]{0.11\\[-0.75ex]$\pm$0.05} & \makecell[c]{6211.46\\[-0.75ex]$\pm$41.25}& \makecell[c]{0.07\\[-0.75ex]$\pm$0.03} & \makecell[c]{6117.48\\[-0.75ex]$\pm$35.74} \\
    	\cline{1-12}
        
        GIN~\cite{jhun2023prediction}  & \makecell[c]{0.11\\[-0.75ex]$\pm$0.06} & \makecell[c]{0.25\\[-0.75ex]$\pm$0.12} & \makecell[c]{0.15\\[-0.75ex]$\pm$0.08} & \makecell[c]{0.56\\[-0.75ex]$\pm$0.10} & \makecell[c]{5412.78\\[-0.75ex]$\pm$33.64}& \makecell[c]{0.16\\[-0.75ex]$\pm$0.08} & \makecell[c]{5574.78\\[-0.75ex]$\pm$36.78}& \makecell[c]{0.14\\[-0.75ex]$\pm$0.07} & \makecell[c]{5407.69\\[-0.75ex]$\pm$41.57}& \makecell[c]{0.13\\[-0.75ex]$\pm$0.06} & \makecell[c]{5398.14\\[-0.75ex]$\pm$39.45}\\
    	\cline{1-12}
        GraphTransformer~\cite{varbella2023geometric}  & \makecell[c]{0.21\\[-0.75ex]$\pm$0.09} & \makecell[c]{0.51\\[-0.75ex]$\pm$0.17} & \makecell[c]{0.30\\[-0.75ex]$\pm$0.11} & \makecell[c]{0.59\\[-0.75ex]$\pm$0.09} & \makecell[c]{4512.97\\[-0.75ex]$\pm$41.05}& \makecell[c]{0.32\\[-0.75ex]$\pm$0.08} & \makecell[c]{4414.86\\[-0.75ex]$\pm$37.56}& \makecell[c]{0.27\\[-0.75ex]$\pm$0.07} & \makecell[c]{4617.99\\[-0.75ex]$\pm$41.27}& \makecell[c]{0.31\\[-0.75ex]$\pm$0.09} & \makecell[c]{4557.63\\[-0.75ex]$\pm$37.68} \\
        \cline{1-12}
        RGCN~\cite{schlichtkrull2018modeling}  & \makecell[c]{0.13\\[-0.75ex]$\pm$0.07} & \makecell[c]{0.43\\[-0.75ex]$\pm$0.15} & \makecell[c]{0.19\\[-0.75ex]$\pm$0.12} & \makecell[c]{0.57\\[-0.75ex]$\pm$0.07} & \makecell[c]{5016.28\\[-0.75ex]$\pm$31.25}& \makecell[c]{0.21\\[-0.75ex]$\pm$0.09} & \makecell[c]{4973.68\\[-0.75ex]$\pm$33.41}& \makecell[c]{0.17\\[-0.75ex]$\pm$0.07} & \makecell[c]{5073.65\\[-0.75ex]$\pm$42.51}& \makecell[c]{0.18\\[-0.75ex]$\pm$0.09} & \makecell[c]{5123.46\\[-0.75ex]$\pm$31.54} \\
    	\cline{1-12}

        HGT~\cite{hu2020heterogeneous}  & \makecell[c]{0.57\\[-0.75ex]$\pm$0.07} & \makecell[c]{0.56\\[-0.75ex]$\pm$0.09} & \makecell[c]{0.56\\[-0.75ex]$\pm$0.09} & \makecell[c]{0.58\\[-0.75ex]$\pm$0.07} & \makecell[c]{2566.34\\[-0.75ex]$\pm$34.25}& \makecell[c]{0.54\\[-0.75ex]$\pm$0.09} & \makecell[c]{2876.89\\[-0.75ex]$\pm$31.54}& \makecell[c]{0.58\\[-0.75ex]$\pm$0.07} & \makecell[c]{2431.78\\[-0.75ex]$\pm$29.84}& \makecell[c]{0.59\\[-0.75ex]$\pm$0.09} & \makecell[c]{3128.39\\[-0.75ex]$\pm$33.66}\\
    	\cline{1-12}

        HGSL~\cite{zhao2021heterogeneous}  & \makecell[c]{0.59\\[-0.75ex]$\pm$0.09} & \makecell[c]{0.65\\[-0.75ex]$\pm$0.12} & \makecell[c]{0.62\\[-0.75ex]$\pm$0.11} & \makecell[c]{0.66\\[-0.75ex]$\pm$0.13} & \makecell[c]{2298.77\\[-0.75ex]$\pm$31.78}& \makecell[c]{0.60\\[-0.75ex]$\pm$0.14} & \makecell[c]{2356.64\\[-0.75ex]$\pm$33.24}& \makecell[c]{0.62\\[-0.75ex]$\pm$0.16} & \makecell[c]{2097.12\\[-0.75ex]$\pm$31.47}& \makecell[c]{0.63\\[-0.75ex]$\pm$0.11} & \makecell[c]{2439.67\\[-0.75ex]$\pm$34.18}\\
    	\cline{1-12}
     
        HINT~\cite{sturaro2018realistic}  & \makecell[c]{0.61\\[-0.75ex]$\pm$0.05} & \makecell[c]{0.72\\[-0.75ex]$\pm$0.04} & \makecell[c]{0.66\\[-0.75ex]$\pm$0.05} & \makecell[c]{0.72\\[-0.75ex]$\pm$0.07} & \makecell[c]{1647.91\\[-0.75ex]$\pm$15.24}& \makecell[c]{0.69\\[-0.75ex]$\pm$0.06} & \makecell[c]{1672.83\\[-0.75ex]$\pm$14.25}& \makecell[c]{$/$} & \makecell[c]{$/$}& \makecell[c]{0.57\\[-0.75ex]$\pm$0.03} & \makecell[c]{1635.31\\[-0.75ex]$\pm$12.76}\\
    	\cline{1-12}

        MHGCN~\cite{yu2022multiplex}  & \makecell[c]{0.63\\[-0.75ex]$\pm$0.07} & \makecell[c]{0.75\\[-0.75ex]$\pm$0.09} & \makecell[c]{0.68\\[-0.75ex]$\pm$0.09} & \makecell[c]{0.76\\[-0.75ex]$\pm$0.07} & \makecell[c]{1264.78\\[-0.75ex]$\pm$22.41}& \makecell[c]{0.61\\[-0.75ex]$\pm$0.09} & \makecell[c]{1376.28\\[-0.75ex]$\pm$26.41}& \makecell[c]{0.62\\[-0.75ex]$\pm$0.07} & \makecell[c]{1222.98\\[-0.75ex]$\pm$25.47}& \makecell[c]{0.65\\[-0.75ex]$\pm$0.09} & \makecell[c]{1421.97\\[-0.75ex]$\pm$31.42}\\
    	\cline{1-12}
     
        \textbf{$I^3$}  & \makecell[c]{\textbf{0.72}\\[-0.75ex]\textbf{$\pm$0.08}} & \makecell[c]{\textbf{0.93}\\[-0.75ex]$\pm$\textbf{0.03}} & \makecell[c]{\textbf{0.81}\\[-0.75ex]$\pm$\textbf{0.04}} & \makecell[c]{\textbf{0.95}\\[-0.75ex]$\pm$\textbf{0.03}} & \makecell[c]{\textbf{981.35}\\[-0.75ex]$\pm$\textbf{31.41}}& \makecell[c]{\textbf{0.82}\\[-0.75ex]$\pm$\textbf{0.04}} & \makecell[c]{\textbf{1022.63}\\[-0.75ex]$\pm$\textbf{24.51}}& \makecell[c]{\textbf{0.84}\\[-0.75ex]$\pm$\textbf{0.03}} & \makecell[c]{\textbf{976.41}\\[-0.75ex]$\pm$\textbf{22.41}}& \makecell[c]{\textbf{0.77}\\[-0.75ex]$\pm$\textbf{0.05}} & \makecell[c]{\textbf{984.66}\\[-0.75ex]$\pm$\textbf{29.64}}\\
    	\cline{1-12}
    	\hline
	\end{tabular}
    \label{tab:prediction}
\end{table*}

\begin{table*}
\renewcommand\arraystretch{0.99}
    \centering
    \tiny
	\caption{
       CF prediction performance evaluation results on interdependent networks in \textbf{REAL-WORLD} dataset. The backslash indicates that the baseline cannot be applied to the analysis of the network. The best results are highlighted in bold.}
	\begin{tabular}{l||c|c|c|c||c||c|c||c|c||c|c}
    	\hline
        \multirow{3}{*}{} & \multicolumn{5}{c||}{$G$}& \multicolumn{2}{c||}{$G_{e}$} & \multicolumn{2}{c||}{$G_{r}$}  & \multicolumn{2}{c}{  $G_{c}$}    \\ \hline
        & Pre & Rec & F1 & AUC & RMSE& F1& RMSE& F1& RMSE& F1& RMSE \\ \hline
        ICM~\cite{goldenberg2001talk}  & \makecell[c]{0.24\\[-0.75ex]$\pm$0.07} & \makecell[c]{0.41\\[-0.75ex]$\pm$0.09} & \makecell[c]{0.30\\[-0.75ex]$\pm$0.08} & \makecell[c]{0.59\\[-0.75ex]$\pm$0.11} & \makecell[c]{5129.64\\[-0.75ex]$\pm$17.66}& \makecell[c]{0.24\\[-0.75ex]$\pm$0.06} & \makecell[c]{5872.09\\[-0.75ex]$\pm$15.19}& \makecell[c]{0.30\\[-0.75ex]$\pm$0.09} & \makecell[c]{5326.88\\[-0.75ex]$\pm$14.99}& \makecell[c]{0.33\\[-0.75ex]$\pm$0.07} & \makecell[c]{5209.87\\[-0.75ex]$\pm$15.44} \\
    	\cline{1-12}
        GCN~\cite{kipf2016semi}  & \makecell[c]{0.22\\[-0.75ex]$\pm$0.06} & \makecell[c]{0.37\\[-0.75ex]$\pm$0.11} & \makecell[c]{0.28\\[-0.75ex]$\pm$0.10} & \makecell[c]{0.57\\[-0.75ex]$\pm$0.13} & \makecell[c]{5328.14\\[-0.75ex]$\pm$36.77}& \makecell[c]{0.25\\[-0.75ex]$\pm$0.09} & \makecell[c]{5610.93\\[-0.75ex]$\pm$37.43}& \makecell[c]{0.28\\[-0.75ex]$\pm$0.13} & \makecell[c]{5484.13\\[-0.75ex]$\pm$37.11}& \makecell[c]{0.22\\[-0.75ex]$\pm$0.07} & \makecell[c]{5981.42\\[-0.75ex]$\pm$41.76} \\
    	\cline{1-12}
        
        GIN~\cite{jhun2023prediction}  & \makecell[c]{0.31\\[-0.75ex]$\pm$0.14} & \makecell[c]{0.41\\[-0.75ex]$\pm$0.16} & \makecell[c]{0.35\\[-0.75ex]$\pm$0.14} & \makecell[c]{0.62\\[-0.75ex]$\pm$0.13} & \makecell[c]{4871.63\\[-0.75ex]$\pm$37.66}& \makecell[c]{0.31\\[-0.75ex]$\pm$0.11} & \makecell[c]{5124.29\\[-0.75ex]$\pm$37.91}& \makecell[c]{0.33\\[-0.75ex]$\pm$0.12} & \makecell[c]{4875.35\\[-0.75ex]$\pm$40.08}& \makecell[c]{0.38\\[-0.75ex]$\pm$0.10} & \makecell[c]{4341.54\\[-0.75ex]$\pm$38.90}\\
    	\cline{1-12}
        GraphTransformer~\cite{varbella2023geometric}  & \makecell[c]{0.58\\[-0.75ex]$\pm$0.13} & \makecell[c]{0.69\\[-0.75ex]$\pm$0.18} & \makecell[c]{0.63\\[-0.75ex]$\pm$0.15} & \makecell[c]{0.69\\[-0.75ex]$\pm$0.17} & \makecell[c]{2763.98\\[-0.75ex]$\pm$42.77}& \makecell[c]{0.61\\[-0.75ex]$\pm$0.18} & \makecell[c]{3068.15\\[-0.75ex]$\pm$47.89}& \makecell[c]{0.53\\[-0.75ex]$\pm$0.14} & \makecell[c]{4017.65\\[-0.75ex]$\pm$45.88}& \makecell[c]{0.63\\[-0.75ex]$\pm$0.16} & \makecell[c]{2965.41\\[-0.75ex]$\pm$45.73} \\
        \cline{1-12}
        RGCN~\cite{schlichtkrull2018modeling}  & \makecell[c]{0.34\\[-0.75ex]$\pm$0.13} & \makecell[c]{0.49\\[-0.75ex]$\pm$0.15} & \makecell[c]{0.40\\[-0.75ex]$\pm$0.14} & \makecell[c]{0.63\\[-0.75ex]$\pm$0.15} & \makecell[c]{4653.22\\[-0.75ex]$\pm$35.66}& \makecell[c]{0.38\\[-0.75ex]$\pm$0.14} & \makecell[c]{4870.42\\[-0.75ex]$\pm$37.87}& \makecell[c]{0.43\\[-0.75ex]$\pm$0.18} & \makecell[c]{4132.70\\[-0.75ex]$\pm$48.37}& \makecell[c]{0.45\\[-0.75ex]$\pm$0.17} & \makecell[c]{3876.08\\[-0.75ex]$\pm$41.02} \\
    	\cline{1-12}

        HGT~\cite{hu2020heterogeneous}  & \makecell[c]{0.69\\[-0.75ex]$\pm$0.17} & \makecell[c]{0.72\\[-0.75ex]$\pm$0.15} & \makecell[c]{0.70\\[-0.75ex]$\pm$0.16} & \makecell[c]{0.78\\[-0.75ex]$\pm$0.14} & \makecell[c]{1689.03\\[-0.75ex]$\pm$31.92}& \makecell[c]{0.72\\[-0.75ex]$\pm$0.17} & \makecell[c]{1498.30\\[-0.75ex]$\pm$34.76}& \makecell[c]{0.67\\[-0.75ex]$\pm$0.18} & \makecell[c]{1987.43\\[-0.75ex]$\pm$39.72}& \makecell[c]{0.74\\[-0.75ex]$\pm$0.19} & \makecell[c]{1347.86\\[-0.75ex]$\pm$31.27}\\
    	\cline{1-12}

        HGSL~\cite{zhao2021heterogeneous}  & \makecell[c]{0.71\\[-0.75ex]$\pm$0.16} & \makecell[c]{0.76\\[-0.75ex]$\pm$0.16} & \makecell[c]{0.73\\[-0.75ex]$\pm$0.15} & \makecell[c]{0.81\\[-0.75ex]$\pm$0.17} & \makecell[c]{1432.65\\[-0.75ex]$\pm$32.02}& \makecell[c]{0.76\\[-0.75ex]$\pm$0.14} & \makecell[c]{1248.97\\[-0.75ex]$\pm$35.03}& \makecell[c]{0.71\\[-0.75ex]$\pm$0.15} & \makecell[c]{1652.38\\[-0.75ex]$\pm$39.44}& \makecell[c]{0.77\\[-0.75ex]$\pm$0.11} & \makecell[c]{1212.31\\[-0.75ex]$\pm$39.11}\\
    	\cline{1-12}
     
        HINT~\cite{sturaro2018realistic}  & \makecell[c]{0.72\\[-0.75ex]$\pm$0.09} & \makecell[c]{0.79\\[-0.75ex]$\pm$0.11} & \makecell[c]{0.75\\[-0.75ex]$\pm$0.11} & \makecell[c]{0.86\\[-0.75ex]$\pm$0.12} & \makecell[c]{1298.33\\[-0.75ex]$\pm$14.08}& \makecell[c]{0.77\\[-0.75ex]$\pm$0.08} & \makecell[c]{1432.18\\[-0.75ex]$\pm$18.49}& \makecell[c]{$/$} & \makecell[c]{$/$}& \makecell[c]{0.73\\[-0.75ex]$\pm$0.09} & \makecell[c]{1273.99\\[-0.75ex]$\pm$14.30}\\
    	\cline{1-12}

        MHGCN~\cite{yu2022multiplex}  & \makecell[c]{0.76\\[-0.75ex]$\pm$0.14} & \makecell[c]{0.84\\[-0.75ex]$\pm$0.17} & \makecell[c]{0.80\\[-0.75ex]$\pm$0.16} & \makecell[c]{0.92\\[-0.75ex]$\pm$0.16} & \makecell[c]{874.38\\[-0.75ex]$\pm$29.68}& \makecell[c]{0.79\\[-0.75ex]$\pm$0.19} & \makecell[c]{964.91\\[-0.75ex]$\pm$29.11}& \makecell[c]{0.75\\[-0.75ex]$\pm$0.15} & \makecell[c]{1134.19\\[-0.75ex]$\pm$22.33}& \makecell[c]{0.82\\[-0.75ex]$\pm$0.14} & \makecell[c]{871.18\\[-0.75ex]$\pm$29.04}\\
    	\cline{1-12}
     
        \textbf{$I^3$}  & \makecell[c]{\textbf{0.83}\\[-0.75ex]\textbf{$\pm$0.13}} & \makecell[c]{\textbf{0.94}\\[-0.75ex]$\pm$\textbf{0.16}} & \makecell[c]{\textbf{0.88}\\[-0.75ex]$\pm$\textbf{0.15}} & \makecell[c]{\textbf{0.97}\\[-0.75ex]$\pm$\textbf{0.18}} & \makecell[c]{\textbf{632.98}\\[-0.75ex]$\pm$\textbf{32.77}}& \makecell[c]{\textbf{0.86}\\[-0.75ex]$\pm$\textbf{0.13}} & \makecell[c]{\textbf{864.98}\\[-0.75ex]$\pm$\textbf{22.03}}& \makecell[c]{\textbf{0.84}\\[-0.75ex]$\pm$\textbf{0.15}} & \makecell[c]{\textbf{913.90}\\[-0.75ex]$\pm$\textbf{20.83}}& \makecell[c]{\textbf{0.89}\\[-0.75ex]$\pm$\textbf{0.07}} & \makecell[c]{\textbf{732.98}\\[-0.75ex]$\pm$\textbf{38.33}}\\
    	\cline{1-12}
    	\hline
	\end{tabular}
    \label{tab:prediction_realworld}
\end{table*}

\subsection{Baselines}
For the CF prediction, we use various baselines ranging from homogeneous to heterogeneous graphs to measure the performance. For homogeneous graphs, we consider the following baselines:
\begin{itemize}[]
\item[$\bullet$]\textbf{ICM}~\cite{goldenberg2001talk}: Each node has a certain probability to propagate its state to its neighboring nodes. The propagation events are independent of each other.
\item[$\bullet$]\textbf{GCN}~\cite{kipf2016semi}: A type of GNN that achieves the extraction and learning of node features by performing convolution operations on the neighbor matrix.
\item[$\bullet$]\textbf{GIN}~\cite{jhun2023prediction}: Fitting cascade propagation under the ML mechanism using a multilayer GIN.
\item[$\bullet$]\textbf{GraphTransformer}~\cite{varbella2023geometric}: A structure capturing CFs in electric networks mainly consisting of three GraphTransformer layers.
\end{itemize}
When using baselines on isomorphic maps, we treated infrastructure networks that are supposed to be heterogeneous graphs as homogeneous graphs. In addition, we also consider baselines on heterogeneous graphs including:
\begin{itemize}[]
\item[$\bullet$]\textbf{RGCN}~\cite{schlichtkrull2018modeling}: A deep learning model for processing graph data with multiple relationships, which updates the feature representation of nodes by aggregating different types of relationships.
\item[$\bullet$]\textbf{HGT}~\cite{hu2020heterogeneous}: A Heterogeneous graph transformer capable of modeling graph heterogeneity. It designs node and edge type dependent parameters to characterize heterogeneous attention on each edge, enabling HGT to maintain dedicated representations for different types of nodes and edges.
\item[$\bullet$]\textbf{HGSL}~\cite{zhao2021heterogeneous}: A Heterogeneous graph Neural Network that considers not only feature similarity by generating feature similarity graphs, but also complex heterogeneous interactions in features and semantics by generating feature propagation graphs and semantic graphs.
\item[$\bullet$]\textbf{HINT}~\cite{sturaro2018realistic}: A model-based model whose core idea is to model different characteristics of nodes by dividing them into different roles in the network.
\item[$\bullet$]\textbf{MHGCN}~\cite{yu2022multiplex}: A Multiple Heterogeneous graph convolutional Network, through multi-layer convolutional aggregation, MHGCN can automatically learn useful heterogeneous meta-path interactions of different lengths in multiple heterogeneous networks. 
\end{itemize}
The baseline we use is comprehensive and representative, including state-of-art on both homogeneous and heterogeneous graphs. In addition, to validate the effectiveness of our components, we constructed the following partial variant model:
\begin{itemize}[]
\item[$\bullet$]\textbf{$I^3$ w/o $E_{lp}$}: $I^3$ removing the link prediction embedding.
\item[$\bullet$]\textbf{$I^3$ w/o $E_p$}: $I^3$ removing the pooling module embedding.
\item[$\bullet$]\textbf{$I^3$ w/o $E_{ie}$}: $I^3$ removing initial node enhanced embedding.
\item[$\bullet$]\textbf{$I^3$ w/o RGCN}: A homogeneous graph neural network with a similar structure to $I^3$, replacing the RGCN in $I^3$ with GCN. 
\end{itemize}

\begin{figure} 
	\centering  
    \includegraphics[width=0.5\linewidth]{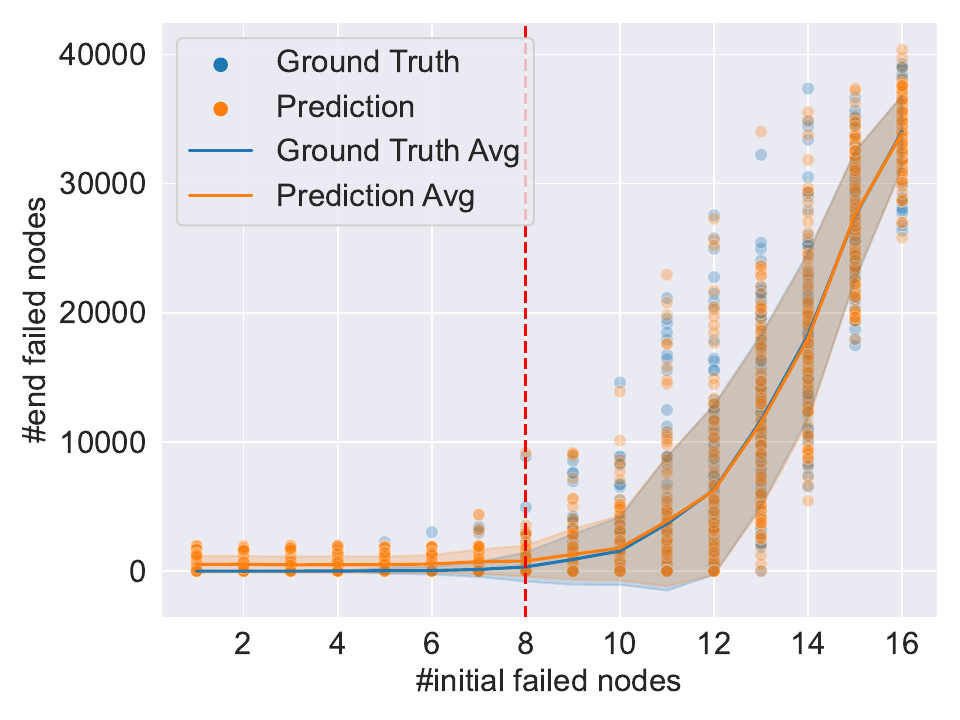}
    \caption{Experiment of Phase Transition}
    \label{Exp4}
\end{figure}

\begin{figure*} 
	\centering  
	\subfigtopskip=2pt 
	\subfigbottomskip=2pt 
	\subfigcapskip=-5pt 
	\subfigure[Total Power of Electric Network]{
		\label{Exp5-Elec}
		\includegraphics[width=0.32\linewidth]{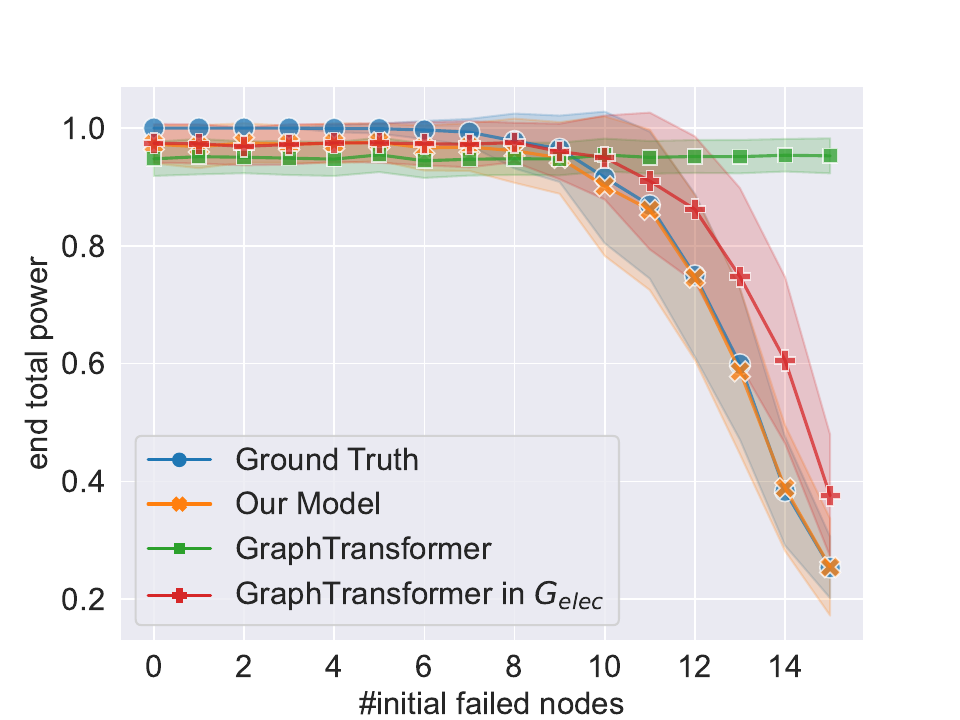}}
    \subfigure[ANC of Road Network]{
		\label{Exp5-Road}
		\includegraphics[width=0.32\linewidth]{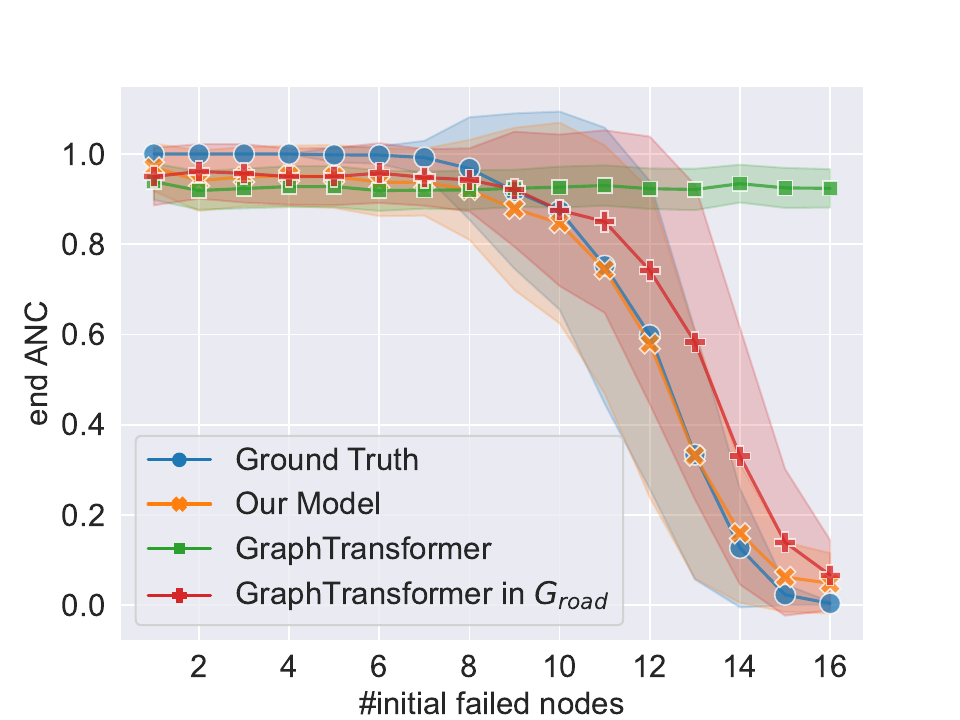}}
    \subfigure[Yield of Communication Network]{
		\label{Exp5-Communication}
		\includegraphics[width=0.32\linewidth]{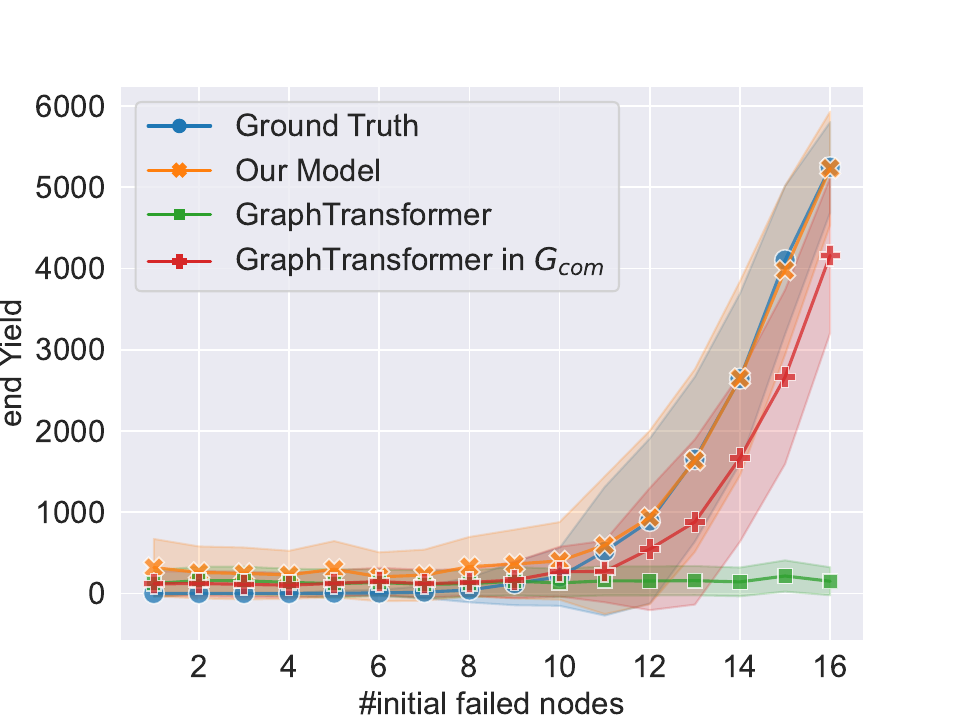}}
    \caption{Experiments of Different Single Networks}
	\label{Exp5}
\end{figure*}

\subsection{Overall Performance (RQ1)}
To examine the performance of our model, we compare it with baselines. The results are shown in Table~\ref{tab:prediction} and Table~\ref{tab:prediction_realworld}. It shows that the prediction performance of the baseline on the heterogeneous graph is much higher than that of the baseline on the homogeneous graph, and our model achieves the best on all the metrics. Specifically, $I^3$ achieved a 31.94\% improvement in terms of AUC, 18.03\% in terms of Precision, 29.17\% in terms of Recall, 22.73\% in terms of F1-score, and 28.52\% improvement in terms of RMSE for the CF volume prediction.

\subsection{Ablation Study (RQ2)}
In order to verify the validity and necessity of the three embeddings and the heterogeneous graph structure we designed, we constructed four partial variants of the model to make predictions under the same experimental settings and analyzed the prediction results, which are shown in Table~\ref{tab:ablation study}. The model without link prediction embedding leads to a reduction in prediction performance because it does not capture the similarity and correlation between nodes. The absence of the pooling module embedding in the model results in suboptimal performance as the global properties are not captured. Due to the lack of capture for the first nodes, the model lacking the initial node enhancement embedding performs poorly and causes over-smoothing in GCN. For the model without heterogeneous graph structures gets the worst performance because it cannot capture the different features of different infrastructures.
\begin{table}
\renewcommand\arraystretch{0.99}
    \centering
	\caption{
       Ablation study of $I^3$ in synthetic dataset. The best results are highlighted in bold.}
	\begin{tabular}{l||c|c|c|c||c}
    	\hline
        & Pre & Rec & F1 & AUC & RMSE \\ \hline
        $I^3$ w/o $E_{lp}$  & {0.54} & {0.71} & {0.61} & {0.86} & {1124.96} \\
    	\cline{1-6}
        $I^3$ w/o $E_{p}$  & {0.68} & {0.74} & {0.71} & {0.91} & {1191.80} \\
    	\cline{1-6}
        $I^3$ w/o $E_{ie}$ & {0.13} & {0.54} & {0.20} & {0.61} & {4157.68} \\
    	\cline{1-6}
        $I^3$ w/o RGCN & {0.08} & {0.22} & {0.12} & {0.54} & {5688.69} \\
    	\cline{1-6}
        \textbf{$I^3$}  & \textbf{0.72} & \textbf{0.93} & \textbf{0.81} & \textbf{0.95} & \textbf{981.35}\\
    	\cline{1-6}
    	\hline
	\end{tabular}
    \label{tab:ablation study}
\end{table}

In summary, all three embeddings and heterogeneous graph structures in our model effectively capture the different characteristics of CF propagation in infrastructure networks.

\subsection{Case Study in Different Cascades (RQ3)}
By analyzing the dataset, we find that the number of eventually failed nodes is not the same for different combinations of initial failed nodes. In other words, the size of the final CF caused by the same two initially failed nodes is not the same because of the different topology and nature of the initially failed infrastructure in the network. Thus, we wonder if $I^3$ can accurately predict CFs for all such different combinations of nodes. Specifically, we draw a heat map of the relationship between different initial number of failed nodes and final failed nodes for the difference between the ground-truth and the prediction of HINT and prediction of $I^3$, respectively. By observing the distribution of the two heat maps, we can visually express the prediction accuracy of $I^3$ for different cascade sizes resulting from different combinations of nodes. The two heat maps we plotted are shown in Fig.~\ref{Exp3}. By observing the heat map we can find that comparing to HINT, $I^3$ is able to better capture the different sizes of cascades caused by different initial failed node combinations. This is because our model introduces the initial node enhancement pre-training task, which can identify different initial failed nodes more efficiently. In addition, our model is also trained based on the dataset and has better robustness compared to the model-based model HINT.

\subsection{Phase Transition in Interdependent Networks (RQ4)}
Previous studies~\cite{duan2019universal,su2014robustness} have shown that there is a phase transition of CFs in infrastructure networks, that is, as the number of failed nodes in the network increases, at first the CFs only propagate locally, but when the number of failed components reaches a certain threshold, the failed cascade suddenly starts to propagate globally. Thus, we wonder if $I^3$ also have the ability to capture and accurately predict phase transitions when they exist in an infrastructure network. Specifically, we restricted the initial failed nodes to only power plants and the result is shown in Fig.~\ref{Exp4}, in which phase transition points are indicated by red dashed lines. We predict the cascade for different initial number of failed nodes and compare the results with the true values. From the results we can find that $I^3$ captures the phase transition points in the infrastructure network well. This is because our model is designed with a global pooling pre-training task to capture higher-order dynamics and global evolution, which in combination with the initial node enhancement pre-training task can discriminate whether cascading failures in the given initial failed node case will propagate locally or globally. This capability allows $I^3$ to model the phase transition point in infrastructure networks.

\subsection{Phase Transition in Single Networks (RQ5)}
In the previous studies, most of the methods focus on CF prediction in single networks, ignoring the role of other network components and therefore not being able to achieve accurate prediction. But $I^3$ is a model for modeling CFs in interdependent networks, which overcomes this shortcoming very well. Thus, it is necessary for us to verify whether $I^3$ can also accurately predict CF on different single networks. Specifically, we train and predict on three models, i.e., $I^3$, homogeneous network with all infrastructures, and homogeneous network with only one type of infrastructure, respectively. We observe the prediction results for CF on the electric network, road network and communication network, respectively. For the cascading failure metric, we have different designs for different networks. For the electric network, we choose the total power of the network. For the transportation network, we chose the ANC as a metric, which is calculated by the following formula:
\begin{align}
	ANC(G,F)=\frac{\sigma(G \textbackslash F)}{\sigma(G)}, \label{ANC}
\end{align}
in which $F$ refers to the set of failed nodes and $\sigma$ refers to the connectivity of graph $G$. Connectivity $\sigma$ can be defined as:
\begin{align}
	\sigma(G)=\sum_{C_i \in G}\frac{\epsilon_i(\epsilon_i-1)}{2}, \label{sigma}
\end{align}
in which $C_i$ is the $i^th$ connected component in graph $G$ and $\epsilon_i$ is the size of $C_i$. For communication networks, since often the failure of a small number of communication base stations does not affect the functionality of the network, thus we choose yield as the metric. Specifically, for each AOI, we consider an AOI to be failed (AOI state is 0) only if more than half of the base station it connected to is failed. Otherwise, we consider it as normal (AOI state is 1). For each cascade failure propagation, we count the number of AOIs that fail at the end as a functional indicator of the communication network. The formula of yield is as follows:
\begin{align}
	\text{Yield}(G)=\frac{\sum_{i \in V_{AOI}}s_i}{N_{AOI}},\label{sigma}
\end{align}
in which $N_{AOI}$ refers to the number of base stations,$V_{AOI}$ refers to the set of AOIs and $s_{i}$ refers to the state of AOI $i$. For each network we draw a plot of the true and predicted values of the final network efficiency metrics with different initial numbers of failed nodes and fit a curve, and the results are shown in Fig.~\ref{Exp5}.

The results indicate that $I^3$ is able to predict the phase transition points of network functions in different single networks well. This is because $I^3$ models different infrastructures using heterogeneous graphs making it capable of capturing different single network dynamics. However, for homogeneous networks with all infrastructures, it is not able to accurately capture the phase transitions of the network because it cannot distinguish between different kinds of nodes. This leads to a great amount of noise in the network, making prediction extremely difficult. For homogeneous networks with only one type of infrastructure, although phase transitions can occur, the phase transition points are later than in the real case. This is because homogeneous networks with only one type of infrastructure reduce the noise caused by different infrastructure nodes compared to homogeneous networks with all infrastructures. But modeling of other infrastructure nodes is neglected, causing the lag of phase transition point, which is also similar to the conclusion in~\cite{xie2021detecting}. 
In summary, $I^3$ can better model phase transitions in different networks and can predict the phase transition point more accurately than previous methods.

\section{Conclusion}
In this paper, we propose an \textbf{I}ntegrated \textbf{I}nterdependent \textbf{I}nfrastructure Cascade Failure Model ($I^3$) to capture CF in urban infrastructure networks. The model uses two GAE backbones and extracts different features of the cascades in infrastructure networks by three carefully-designed embeddings. Extensive experiments proved the superiority and robustness of the model. Future work directions can consider incorporating more a prior knowledge in the network such as power flow calculation in the grid, or incorporating timing information for temporal prediction. More interactions between different infrastructure networks could also be applied or even the use of more complex graph structures, such as hypergraphs, to model infrastructure networks. These improvements are important for real-time monitoring and controlling the propagation of cascades.



\bibliographystyle{ACM-Reference-Format}
\bibliography{sample-base}

\end{document}